\documentclass[sigconf]{acmart}
\usepackage{booktabs} 
\usepackage{multirow}
\usepackage{footnote}
\usepackage{tablefootnote}
\makesavenoteenv{tabular}
\usepackage{times}
\usepackage{soul}
\usepackage{url}

\newcommand{\plotECNE}{\addplot[color=violet!90!black, mark=triangle*, mark size=4pt]}
\newcommand{\plotECNED}{\addplot[color=violet!70!black, mark=triangle, mark size=4pt]}

\newcommand{\plotNetMF}{\addplot[color=blue, mark=diamond*, mark size=4pt]}
\newcommand{\plotNodeToVec}{\addplot[color=black, mark=square*, mark size=4pt]}
\newcommand{\plotDeepWalk}{\addplot[color=brown, mark=star, mark size=4pt]}
\newcommand{\plotGraphSage}{\addplot[color=blue!90!black, mark=square, mark size=4pt]}
\newcommand{\plotSDNE}{\addplot[color=green!90!black, mark=star, mark size=4pt]}
\newcommand{\plotstructtovec}{\addplot[color=red, mark=triangle, mark size=4pt]}

\newcommand{\plotECNEH}{\addplot[style={violet!90!black, fill=violet!90!black,mark=none}]}
\newcommand{\plotECNEDH}{\addplot[style={pink, fill=pink,mark=none}]}

\newcommand{\plotNetMFH}{\addplot[style={blue, fill=blue,mark=none}]}
\newcommand{\plotNodeToVecH}{\addplot[style={black, fill=black,mark=none}]}
\newcommand{\plotDeepWalkH}{\addplot[style={brown, fill=brown,mark=none}]}

\newcommand{\plotGraphSageH}{\addplot[style={yellow, fill=yellow, mark=none}]}
\newcommand{\plotSDNEH}{\addplot[style={green, fill=green, mark=none}]}
\newcommand{\plotstructtovecH}{\addplot[style={red, fill=red, mark=none}]}

\newcommand{\scale}{0.57652066}
\newcommand{\spaceMiniPage}{0.2561174}
\newcommand{\scaleClust}{0.93}

\usepackage{pgfplots}
\usepackage{pgfplotstable}
\usetikzlibrary{plotmarks}
\usepackage{pgfgantt}
\usepackage{booktabs} 
\usepackage{multirow}

\newcommand{\entity}[1]{\textsf{#1}}

\newcommand{\emb}[1]{{ \entity{EmbNode}(#1)}}
\newcommand{\embE}[1]{{ \entity{EmbEdge}(#1)}}

\def\addlegendimage{\csname pgfplots@addlegendimage\endcsname}
\newenvironment{customlegend}[1][]{%
	\begingroup
	\csname pgfplots@init@cleared@structures\endcsname
	\pgfplotsset{#1}%
}{%
	\csname pgfplots@createlegend\endcsname
	\endgroup
}%
\pgfplotsset{compat=1.13}
\definecolor{Gray}{gray}{0.95}
\definecolor{LGray}{gray}{0.75}
\definecolor{bblue}{HTML}{4F81BD}
\definecolor{rred}{HTML}{C0504D}
\definecolor{ggreen}{HTML}{9BBB59}
\definecolor{ppurple}{HTML}{9F4C7C}
\usetikzlibrary{shapes.geometric,arrows.meta} 
\tikzset{vertex style/.style={
		draw=#1,
		thick,
		fill=#1!70,
		text=white,
		ellipse,
		minimum width=2cm,
		minimum height=0.75cm,
		font=\small,
		outer sep=0.3pt,
	},
	text style/.style={
		sloped,
		text=black,
		font=\footnotesize,
		above
	}
}
\tikzset{
	vector/.pic={
		\draw (0,0) rectangle (0.5,0.25) rectangle (1,0) rectangle (1.5,0.25) rectangle (2,0); 
	}
}

\usepackage{color}

\newtheorem{defn}[theorem]{\normalfont  \textbf{Definition}}

\newcommand{\kg}{{G}}
\newcommand{\kgDef}{\kg=({V}_{G},{E}_{G})}
\newcommand{\lkg}{\mathcal{G}_{L}}

\usepackage{amsmath}
\usepackage{graphicx}
\usepackage{algorithm2e}
\usepackage{algorithmic}    
\usepackage{amsfonts}
\usepackage{cleveref}
\crefformat{section}{\S#2#1#3}
\crefname{section}{§}{§§}
\Crefname{section}{§}{§§}
\usepackage{pgfplots}
\usepackage{pgfplotstable}
\newcommand{\system}{\textsf{ECNE}}
\newcommand{\systemD}{\textsf{ECNE}$^\textbf{d}$}
\newcommand{\systemLP}{\textsf{ECNE-LP}}

\title{Toward Edge-Centric Network Embeddings}

\author{
Giuseppe Pirr\`o \\ Sapienza University of Rome, Italy \\ pirro@di.uniroma1.it
}

\hypersetup{draft}

\acmConference[Draft]{y}{Nov}{2020} 
\acmYear{2020}
\copyrightyear{2020}
\acmPrice{00.00}

\begin{document}

\begin{abstract}
Existing network embedding approaches tackle the problem of learning low-dimensional node representations. However, networks can also be seen in the light of edges interlinking pairs of nodes. The broad goal of this paper is to introduce edge-centric network embeddings. We present an approach called ECNE, which instead of computing node embeddings directly, computes edge embeddings by relying on the notion of \textit{line graph} coupled with an edge weighting mechanism to preserve the dynamic of the original graph in the line graph. We also present a link prediction framework called ECNE-LP, which given a target link $(u,v)$ first collects paths between nodes $u$ and $v$, then directly embeds the edges in these paths, and finally aggregates them toward predicting the existence of a link. We show that both ECNE and ECNE-LP bring benefit wrt the state-of-the-art.
\end{abstract}

{\tiny }\maketitle
\section{Introduction}
\label{sec:introduction}
The problem of learning network representations, in the form of low-dimensional embeddings, has been extensively studied in the last years \cite{cai2018comprehensive,wu2019comprehensive}. In particular, a variety of approaches has been proposed to \textit{learn node embeddings} such that nodes that are (structurally) similar have embeddings that are close together. Early work like Deepwalk \cite{perozzi2014deepwalk} and node2vec \cite{grover2016node2vec} to define node similarity leverage the notion of context, which mimics on graphs the reasoning behind the SkipGram model used to learn word embeddings from text~\cite{mikolov2013distributed}. 
Here, random walks in a graph (i.e., sequences of neighborhood nodes) play the same role as text sentences (i.e., sequences of words).  Struc2vec \cite{ribeiro2017struc2vec} is a more recent approach also based on random walks. LINE~\cite{tang2015line} guides the generation of random walks by using 1-hop and 2-hop neighborhoods. On the other hand, SDNE \cite{wang2016structural} uses autoencoders to preserve structural similarity. A very recent piece of work is NetMF \cite{qiu2018network}, which bridges the worlds of matrix factorization and random-walk-based approaches.

Other categories of node embedding techniques include Graph Neural Networks (GNNs) that refine the notion of node neighborhood; for instance, GraphSage \cite{hamilton2017inductive} incorporates fixed-size neighborhood information of nodes when computing their embeddings. Graph Convolutions Networks (GCN)~\cite{kipf2017semi} port the convolutions operations on graphs while Graph Attention Networks (GAT) introduce mechanisms to weight the importance of all neighbors \cite{velivckovic2017graph}. Despite the variety of existing approaches, we observe that \textit{all of them have focused on learning node embeddings only, although networks can also be seen in the lights of their edges}. 

\subsection{Motivation}
Edge-centric views of networks have shown their usefulness in a variety of tasks. Evans and Lambiotte \cite{evans2009line} studied how to define communities as a partition of the links rather than the nodes of a network. It has been shown that link communities naturally incorporate overlap while revealing hierarchical organization and that a link-based approach is superior to existing node-based approaches  \cite{ahn2010link} especially when one needs to find overlapping communities. Edge-centric network analysis is very important in biology and spans from the analysis of reaction networks \cite{nacher2004clustering} to the human connectome \cite{de2014edge}, that is, the network resulting from neural interactions.

We observe that embeddings gained popularity much later than the above-mentioned pieces of work \cite{mikolov2013distributed}. {Therefore, we believe that rethinking edge-centric tasks on the light of edge embeddings can provide a refreshing perspective}. We will provide an extensive evaluation on the task of edge-centric community detection (\cref{sec:exp-edge-embeddings}). Furthermore, a number of recent applications making usage of node embeddings can be also rethought of in the light of edge embeddings. As an example, we show how recent approaches for link prediction that leverage paths between a pair of nodes to establish the plausibility of a link between them (e.g., \cite{agrawal2019learning}), can benefit from edge embeddings. With our approach, instead of vectorizing paths as sequences of node embeddings, we can vectorize them as sequences of edge embeddings able to better capture the peculiarity of each link in a path (\cref{sec:link-prediction}).

\subsection{Contributions and Outline}
We set two main goals in this paper. The first is to present a framework to \textit{directly learn edge embeddings instead of node embeddings}. The second is to show the \textit{usefulness of edge embeddings in the task of link prediction} (see \cite{martinez2017survey} for a survey), which is usually tackled by leveraging node embeddings. 

\textit{To tackle the first goal}, we introduce the Edge Centric Network Embedding (\system) technique, which relies on the notion of line graph of a graph \cite{whitney1992congruent}.  The line graph $\lkg$  of a graph $G$ is such that its nodes are edges of the original graph and an edge is inserted between adjacent nodes (i.e., edges sharing a node). 

Once the line graph is available, our approach could benefit from any existing node embedding technique (e.g., node2vec, Deepwalk, sdne, GraphSage) as learning node embeddings of $\lkg$ leads, by construction, to learn the embeddings of the edges of $G$. However, directly working with $\lkg$, leads to low-quality embeddings because edges in the $\lkg$ are added based on adjacency; basically, nodes having a high-degree in $G$ get over-represented in $\lkg$, which does not correctly reflect the dynamics of $G$ in  $\lkg$. To overcome this problem, \system\ introduces an edge weighting strategy for $\lkg$ based on the centrality of nodes of $G$.  However, the size of the line graph is usually much larger than that of the original graph. To overcome this issue, in the implementation of \system\ we make usage of a recent graph coarsening strategy \cite{liang2018mile}.  The idea is to repeatedly coarsen the (line) graph into smaller ones, apply any embedding method on the coarsest graph and refine the embeddings to the original graph through a graph convolution neural network. We show that \system's \textit{direct} edge embedding approach can scale to large graphs and brings an improvement as compared to the indirect way based on the aggregation of edges endpoints' embeddings (\cref{sec:exp-edge-embeddings}). 

\textit{To tackle the second goal}, we present a learning model called \systemLP, which starts by collecting paths between the nodes $u$ and $v$ for which the existence of a link has to be estimated. Then, it embeds the edges in these paths via \system, and finally aggregates them, according to different strategies, toward providing a verdict for the $(u,v)$ link. 

\medskip
\noindent
The main contributions of this paper are as follows: 
\begin{enumerate}
	\item A direct way to compute edge embeddings based on the notion of line graph;
	\item an edge-weighting mechanism for the line graph, which preserves the dynamics of the; original graph in the line graph;
	\item an approach for link prediction based on path embeddings;
	\item an extensive experimental evaluation and comparison with related work.
\end{enumerate}

\medskip
The remainder of the paper is organized as follows. We provide some definitions (\cref{sec:preliminaries}). Then, introduce \system\ (\cref{sec:edge-embeddings}) and the \systemLP\ link prediction approach (\cref{sec:link-prediction}). We report on an experimental evaluation (\cref{sec:experiments}), draw some conclusions and sketch future work (\cref{sec:conclusions}).
%
%
\section{Preliminaries}
\label{sec:preliminaries}
\system\ works on a given graph $\kgDef$, where $V_G$ is the sets of node and $E_G$$\subseteq V_G \times V_G$ is the set of edges. An edge can be directed or undirected, weighted or unweighted, and signed or unsigned. A path $\pi$ of length $l$ between the pair of nodes $(u,v)$ consists of a sequence of the from $\pi$=($u_0$,$u_1$),...,($u_k$,$u_l$), with $u_0$=$u$ and $u_l$=$v$.
We denote by $\Pi^l_{(u,v)}=\{\pi_1^l,\pi_2^l,...\pi_n^l\}$ the set of paths of length $l$ between $u$ and $v$.
\subsection{Line Graph}
The main intuition behind \system\ is to find a way to turn edges of the original graph into nodes and then apply existing node-embeddings methods. The notion of line graph~\cite{whitney1992congruent} is crucial toward this goal.
\begin{defn}
	Given a graph $\kgDef$, its line graph $\lkg=(V_L,E_L)$ is such that: (i) each node of $\lkg$ represents an edge of $\kg$; (ii) two vertices of $\lkg$ are adjacent if, and only if, their corresponding edges in $\kg$ have a node in common.
\end{defn}

Starting from $\kg=(V_G,E_G)$ it is possible to compute the number of nodes and edges of $\lkg=(V_L,E_L)$ as follows: \textit{(i)} the number of nodes of $\lkg$ is equals to the number of edges of $\kg$, i.e., $|V_L|=|E_G|$; \textit{(ii)} the number of edges is $|E_L| \propto \frac{1}{2} \sum_{v\in V_G} d_v^2 -|E_G|$, where $d_{v}$ denotes the degree of the node $v\in V_G$.
The concept of line graph has been extended to other types of graphs, including multigraphs and directed graphs. 

\smallskip
\noindent
\subsection{Node Embeddings}
Given a graph $\kgDef$ and a predefined dimensionality $d$ ($d$$<<$$|V_G|$), the problem of graph embedding is to learn a $d$-dimension vector representation for each node in $\kg$ that best preserve the properties of $G$. If we see a graph as an adjacency matrix, an embedding is essentially a function  $f_n$: R$^{|V_G|\times|V_G|}$ $\to$ R$^{|V_G|\times d}$, which maps the adjacency matrix to a lower dimension matrix. Motivated by the fact that existing graph embedding methods focus on the embeddings of nodes, our goal is to define a \textit{direct way for computing edge embeddings}. In other words, given a graph $\kgDef$, devise an edge embedding function $f_e$: R$^{|E_G|\times|E_G|}$ $\to$ R$^{|E_G|\times d}$, this time considering edge-adjacency.
\section{\system: Edge-Centric Network Embeddings}
\label{sec:edge-embeddings}
We now outline th edge-centric network embedding approach. The idea is to leverage the line graph of a graph where edges of the original network become nodes. With this in mind, once the line graph of a graph has been constructed, any existing node embedding approach can be used in principle. However, we show that directly working on the line graph is not enough and present a strategy to learn more precise edge embeddings (\cref{sec:edge-weights}). 
%
\subsection{Weighted Line Graph}
\label{sec:edge-weights}
The structure of the line graph $\lkg$ is such that high-degree nodes in the original graph $G$ are over-represented in the line graph; a node of $G$ having degree $k$ creates $k(k-1)/2$ edges in $\lkg$. While the Whitney graph isomorphism \cite{whitney1992congruent} does guarantee that the line graph preserves the topology of the original graph, it does not offer the same guarantee when it comes to the dynamics. This can be better understood in terms of how a random walker walks in $G$ and $\lkg$. 

In general, a random walk on each \textit{node} of $G$ will pass along a generic edge $e$ with a certain frequency $f^e$. As this edge is mapped to node $v_e$ in $\lkg$, the same random walk on $\lkg$ the frequency of visiting $v_e$ can be completely different from $f^{v_e}$. Concretely, if we were to apply approaches like Deepwalk or node2vec to $\lkg$ as it is, we would obtain low-quality embeddings as also confirmed by experiments on several networks in an early-version of \system. 

To overcome this issue, one can think of a weighting mechanism. \cite{evans2009overlappingC} (page 7), proposed a weighted line graph where edges are scaled by a factor $O(1/k)$. In this work, we consider an edge weighting mechanism based on current-flow betweenness~\cite{BrandesF05}, which considers the importance of each node in a graph in terms of the number of times it lies on a path between two other nodes (note that it extends the notion of betweenness centrality which focuses on shortest paths only). This allows to better differentiate the edges in $\lkg$ assigning a value that does not only depend on the degree of a node of $G$. 

In particular, the edge from $n_p$=$(i,j)$ to $n_q$=$(j,k)$ in $\lkg $ (representing a path from $i$ to $k$ passing through $j$ in $\kg$) is assigned a weight as follows:
$$w(n_p, n_q)= 1/cb(i) + 1/cb(j) + 1/cb(k)$$

with $cb(x)$ being the current-flow centrality of the node $x$ in $G$, with $x \in \{i,j,k\}$. We leave the investigation of other weighing mechanisms as future work.
%
\subsection{Computing Edge Embeddings}
\label{sec:computing-walks}
Once the (weighted)  line graph is available, \system\ can learn the final edge embeddings by using a variety of existing techniques (e.g., Deepwalk \cite{perozzi2014deepwalk}, node2vec \cite{grover2016node2vec}, SDNE \cite{wang2016structural}). In the current implementation \system\ uses a language model approach based on a set of truncated random walks $\mathcal{W}$. The embedding \system\ learns is a function $f_e:\mathcal{V}_L\rightarrow {R}^d$, which projects nodes of the weighted line graph $\lkg$ into a low dimensional vector space, where $d \ll |\mathcal{V}_L|$, so that neighboring nodes are close in the vector space.
For every node $u \in \mathcal{V}_L$, $N(u)\subset \mathcal{V}_L$ is the set of neighbors and the co-occurrence probability of two nodes $v_i$ and $v_{i+1}$ in a set of walks $\mathcal{W}$ is given by:
\begin{equation}\label{eq:nextnode}
p((e_{v_i},e_{v_{i+1}})\in \mathcal{W}) =\sigma(e_{v_i}^Te_{v_{i+1}})
\end{equation}

\noindent
where $\sigma$ is the softmax function and $e_{v_i}^Te_{v_{i+1}}$ is the dot product of the vectors $e_{v_i}$ and $e_{v_{i+1}}$
As the computation of (\ref{eq:nextnode}) is demanding~\cite{grover2016node2vec}, we use negative sampling to training the Skip-gram model \cite{mikolov2013distributed}. Negative sampling randomly selects nodes that do not appear together in a walk as negative examples, instead of considering all nodes in a graph. 
\section{Path Embedding for Link Prediction}
\label{sec:link-prediction}
%
We now focus our attention on the link prediction task (see \cite{goyal2018graph} for a survey) and present an approach called \systemLP, which leverages edge embeddings. As the goal of this paper is not to specifically tackle link prediction, but to show the potential usage of edge embeddings in a downstream application, we consider the state-of-the-art LEAP system \cite{agrawal2019learning} and adapt it to use edge embeddings instead of node embeddings.  

\begin{figure*}[!h]
	\centering
	\includegraphics[width=.8\textwidth]{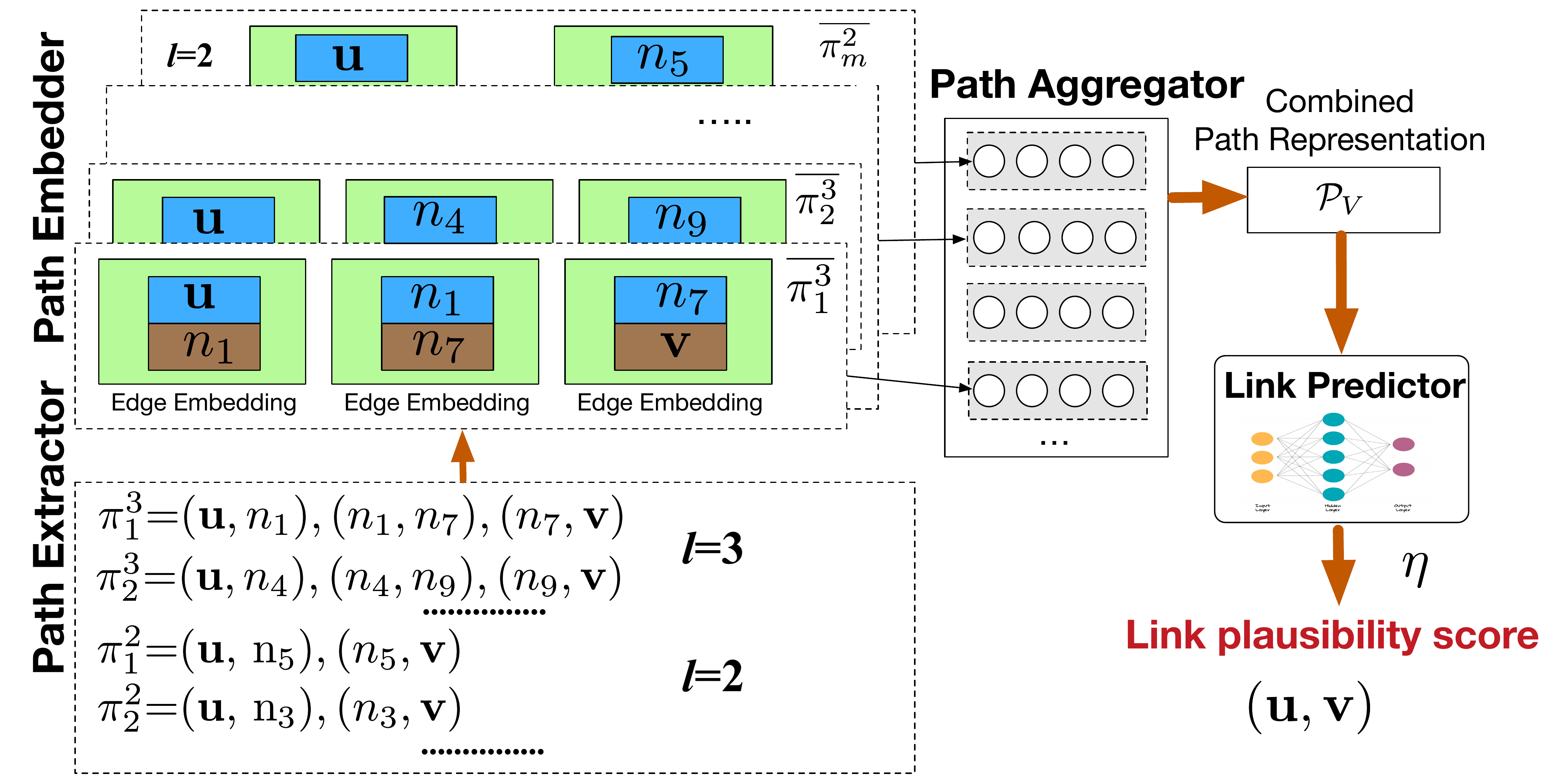}
	\caption{{Overview of \systemLP. Paths are grouped and processed for each length $l$ and then embedded (Path Embedder). An overall path representation is learned (Path Aggregator), which is used to train a binary classifier (Link Predictor).}}
	\label{fig:learning-model}
\end{figure*}

The problem we face can be stated as follows: given a pair of nodes $(u, v)$ assess whether a link between them should hold. The idea is to leverage paths between $u$ and $v$ to collect structural information that can help in assessing the plausibility of a link between $u$ and $v$. \systemLP\ is outlined in Fig.~\ref{fig:learning-model}. It includes three main modules: (i) path extractor; (ii) path embedder; (iii) path aggregator, (iv) link predictor. We will outline each of them in the following.

\subsection{Path Finder}
\label{sec:path-finder}
This module is responsible for finding a set of paths between nodes $u$ and $v$ that will help in assessing the plausibility of the link $(u, v)$. Due to the potential large number of paths, to make this step practically feasible in all networks considered in the experiments (\cref{sec:experiments}), we use the efficient graph-tool library\footnote{\url{https://graph-tool.skewed.de/}} to obtain paths. We note that also LEAP starts from a set of paths between the nodes $u$ and $v$. However, what makes \systemLP\ is the way these paths are embedded  for link prediction.
\subsection{Path Embedder}
\label{sec:path-embedder}
To be processed by the link prediction learning model, paths found by the Path Finder are given a numerical representation. While state-the-art approaches for link prediction, including our direct competitor LEAP \cite{agrawal2019learning}, consider paths as sequences of node embeddings, \systemLP\ considers them as sequences of \textit{edge embeddings}. Consider the paths $\pi_1$=$n_0$-$n_7$-$n_1$ and $\pi_2$=$n_2$-$n_1$-$n_4$-$n_0$. While approaches based on node embeddings will consider in these paths the same embeddings $\emb{n_0}$ and  $\emb{n_1}$ for the nodes $n_0$ and $n_1$, \systemLP\ will consider the specific edge embeddings $\embE{n_0,n_7}$, $\embE{n_4,n_0}$, $\embE{n_7,n_1}$, and $\embE{n_2,n_1}$ for the edges involving these nodes. This will help the learning model to better represent and differentiate paths.
The  function \embE{$\cdot$} used to vectorize each edge in a path  can use either a direct or indirect approach.  The direct  approach is to use \system\ where a path $\pi$=\{$e^1,e^2,\ldots e^l$\} of length $l$ including $l$ edges is encoded as a sequence $\pi_E$=[$e_E^1,e_E^2,\ldots,e_E^l$], where $e_E^i$=\system($e^i$).

As for the indirect way, one can start with node embeddings found by any existing mechanism (e.g., node2vec  \cite{grover2016node2vec}, Deepwalk \cite{perozzi2014deepwalk}, etc) and consider a generic function \emb{$\cdot$}, which given a node, returns its corresponding vector embedding. Hence, to compute the embedding of an edge $e$=$(u,v)$, we can perform some operation $op$ (e.g., concatenation) on its constituents vectors, that is, \embE{$e$}=$op(\emb{u},\emb{v})$. As an example, Grover and Leskovec \cite{grover2016node2vec} considered some operators (e.g., average, Hadamard product) over the embeddings of individual node found via node2vec for link prediction. 
%
%
\subsection{Path Aggregator}
\label{sec:path-aggregator}
Paths converted into their vector form are given an aggregate representation. We see aggregation as a black-box learning module, which takes the {vectorized} paths and provides an overall vector representation for them. Aggregation has been extensively considered in the literature (e.g., \cite{hamilton2017inductive}). In this paper, as our direct competitor is LEAP, we use the  three aggregation strategies used by LEAP.
\subsubsection{Average Pool}
\label{sec:avg-pool-aggr}
This kind of aggregator combines the different representations of paths by concatenating the vector representations of the edges in a path. Then on the set of paths obtained, the aggregator performs a 1D average pooling operation. The final combined path representation is a single vector obtained by averaging the paths between $u$ and $v$ of length $l$. The whole operation can be summarized as follows:
\begin{equation}
\mathcal{P}_V^l={AvgPool([\oplus(\pi_i^l), \forall \pi_i^l \in \mathcal{P}^l])}
\end{equation}

where $AvgPool$ is the one-dimensional average pooling operation, and $\oplus(\cdot)$ is the vector concatenation operation, which concatenating multiple vectors together. This representation relies on the embeddings of the edges in each path.
\subsubsection{Max Pool}
\label{sec:sense-max-aggregator}
This kind of aggregator shares with the AvgPool the fact representation obtained, even in this case, by concatenating edge vectors; what changes is the final vector of the path. Instead of being the average, it is now computed by using a dense neural network layer. The resulting activations are then passed through a max-pooling operation which helps to derive a single vector representation for the paths of length $l$. The whole operation can be summarized as follows:
\begin{equation}
\mathcal{P}_V^l={MaxPool([ \sigma(W_l\cdot\oplus(\pi_i^l)+b_l), \forall \pi_i^l \in \mathcal{P}^l])}
\end{equation}

where $MaxPool$ is the one-dimension max pool operation (which selects bit-wise the maximum value from multiple vectors to derive a single final vector.), $W_l$ are the weights to be learned, $b_l$ the bias, and $\sigma$ the activation function.
\subsubsection{LSTM Max Pool}
\label{sec:recurrent-aggregator}
\begin{figure}[!h]
	\centering
	\includegraphics[width=\columnwidth]{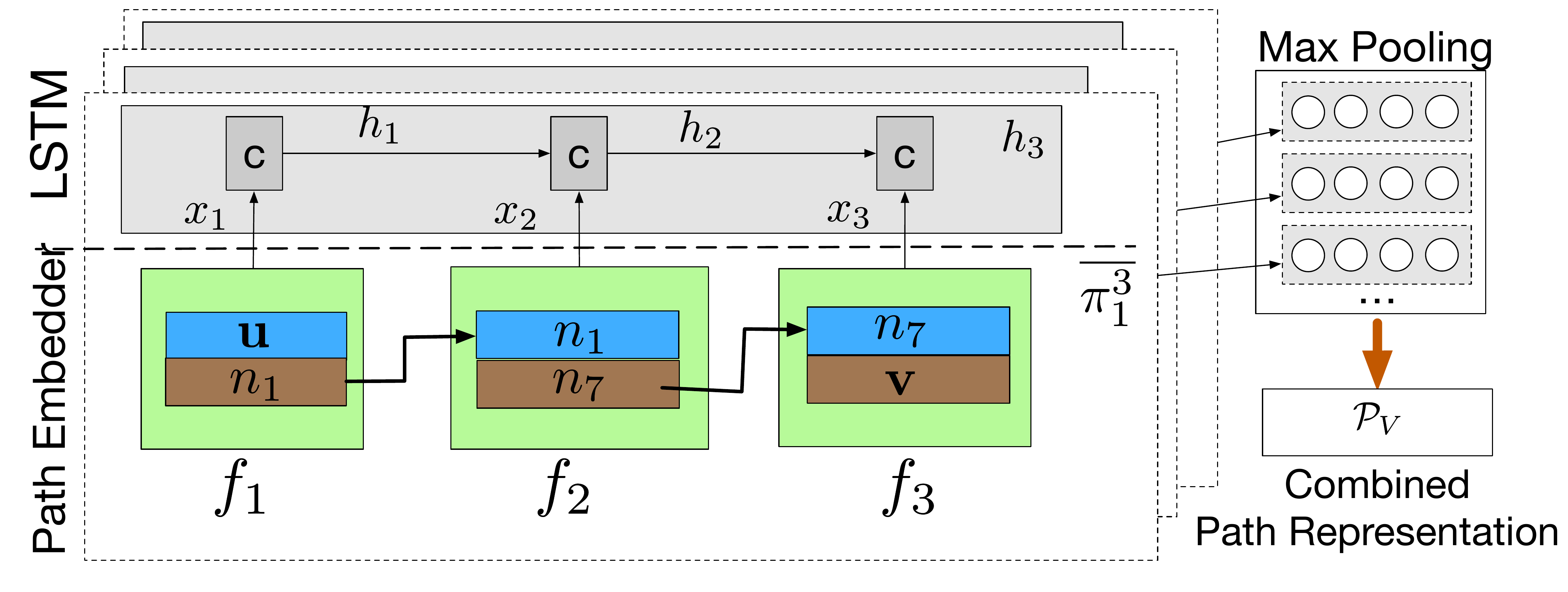}
	\vspace{-.7cm}
	\caption{{An overview of the LSTM Max Pool Aggregator.}}
	\label{fig:lstm-aggregator}
	%
\end{figure}

We now outline the most sophisticated aggregator we considered. The idea is to treat a (vectorized) path as a sequence an employ an LSTM network to cater for sequential dependencies between edges in a path. With this reasoning, each edge in a path represents a point of a sequence. Fig.~\ref{fig:lstm-aggregator} provides an overview of the architecture of the \textit{LSTM Max Pool} aggregator. At each step $l-1$,  the  LSTM layer outputs a hidden state vector $h_{l-1}$, consuming sub-sequence of embedded edges $[f_1,..., f_{l-1}]$. In other words, $x_{l-1}$=$f_{l-1}$. The input $x_{l-1}$ and the hidden state $h_{l-1}$ are used to learn the hidden state of the next path step $l$. 
As our final goal is to leverage the representations of all paths, after processing all of them via the LSTM, the aggregator employs another LSTM followed by a max pool operation (see Fig. \ref{fig:lstm-aggregator}) to produce the combined path representation $\mathcal{P}_V$.

\begin{table*}[!h]
	\caption{Summary of the datasets used for evaluation.}
	\begin{tabular}{|c|c|c|c|}
		\hline
		\textbf{Task}                          & \textbf{Name} & \textbf{Nodes} & \textbf{Edges} \\ \hline
		\multirow{2}{*}{}                      & Karate Club   & 34             & 78             \\ \cline{2-4} 
		& Power Grid    & 4,941          & 6,594          \\ \cline{2-4} 
		\textit{Classif \& Clustering} & Facebook      & 4,039          & 88,234         \\ \cline{2-4} 
		\multirow{3}{*}{}                      & Erdos         & 6,100          & 9,939          \\ \cline{2-4} 
		& Astrophysics  & 17,900         & 197,000        \\ \cline{2-4} 
		& ArXiv         & 18,772         & 198,110        \\ \hline
		& USAir         & 332            & 2,126          \\ \cline{2-4} 
		& NS            & 1,589          & 2,742          \\ \cline{2-4} 
		\textit{Link Prediction}               & PB            & 1,222          & 16,714         \\ \cline{2-4} 
		& C. Ele        & 297            & 2,148          \\ \cline{2-4}
		& E. coli       & 1,805          & 14,660         \\ \cline{2-4}
		& ArXiv         & 18,772         & 198,110        \\ \cline{2-4} 
		& DBLP       & 123,456      & 651,756       \\ \hline
	\end{tabular}
	\label{tab:datasets}
\end{table*}

\subsection{Link Predictor}
\label{sec:fact-checker}
As paths are grouped according to their different lengths, the aggregation mechanism processes each set of paths separately. Finally, the path representations for each length are concatenated together to give the final length-specific path representation $\mathcal{P}_V$ (see Fig. \ref{fig:learning-model}).

The last step consists of providing the final prediction about a link. This is done by the link predictor module, which takes as input the output of the Path Aggregator (i.e., the vector representation $\mathcal{P}_V$) and feeds it into a classifier. We treat the link prediction problem as a binary classification problem, where an existing link and non-existing link are assigned 1 and 0 as target values, respectively. The final goal is to optimize the negative log-likelihood objective function, which defined  as follows:
\begin{equation}
\mathcal{L}=- \sum_{f^+  \in \mathcal{F}^+}   log \ \hat{y}_{f^+} + \sum_{f^-  \in \mathcal{F}^-}log(1- \ \hat{y}_{f^-})
\end{equation}

where $\mathcal{F}^+$=\{$f^+\mid y_{f^+}=1$\} and $\mathcal{F}^-$=\{$f^+\mid y_{f^-}=0$\} are the true and false links, respectively. Given a target link $({u}, {v})$, the link predictor outputs a plausibility score $\eta \in [0,1]$.
\section{Evaluation}
\label{sec:experiments}
The goal of the experimental evaluation was twofold. First, to show that directly computing edge embeddings leads to better performance as compared to indirect edge embeddings obtained from edges' endpoints. Second, to show that edge embeddings are useful in applications where paths between nodes in a network can be exploited. The framework has been implemented in Python using Keras\footnote{\url{http://keras.io}}.

Moreover, to make the computation of (edge) embeddings feasible on the line graph, which usually contains a larger number of nodes, the implementation of \system\ leverages the MILE framework \cite{liang2018mile}. It uses MILE\footnote{\url{http://jiongqianliang.com/MILE/}} to repeatedly coarsen the weighted line graph into smaller ones, apply the embedding method described above on the coarsest graph and refine the embeddings to the original graph via graph convolution. We make available the code of \system\footnote{The code is available upon request.} 

\subsection{Datasets and Experimental Setting}
\label{sec:exp-setting}
We performed experiments to investigate the above-mentioned goals on the real-world datasets summarized in Table~\ref{tab:datasets}. The datasets for edge classification and clustering are popular networks\footnote{We downloaded them from \url{http://snap.stanford.edu}} used in the context of community detection~\cite{grover2016node2vec}.
As for link prediction, we considered the datasets used to evaluate the state-of-the-art LEAP systems~\cite{agrawal2019learning}, which is our direct competitor\footnote{\url{https://github.com/rakshit-agrawal/LEAP}}. 
\begin{figure*}[!h]
	\centering
	\begin{tikzpicture}
	\begin{customlegend}[legend columns=8,legend style={align=left,draw=none,column sep=0ex},legend entries={\system,\systemD,NetMF,node2vec, Deepwalk, struct2vec, sdne, GraphSage}]
	 \addlegendimage{color=violet!90!black, mark=triangle*, mark size=4pt}
		\addlegendimage{color=violet!70!black, mark=triangle, mark size=4pt}
		\addlegendimage{color=blue, mark=diamond*, mark size=4pt}
	\addlegendimage{color=black, mark=square*, mark size=4pt}
	\addlegendimage{color=brown, mark=star, mark size=4pt}
		\addlegendimage{color=red!90!black, mark=triangle, mark size=4pt} 
	   \addlegendimage{color=green, mark=star, mark size=4pt} 
		\addlegendimage{color=blue!90!black, mark=square, mark size=4pt}
	\end{customlegend}
	\end{tikzpicture}
	
	\begin{minipage}{\spaceMiniPage\textwidth}
		\begin{tikzpicture}[scale=\scale][font=\Huge]
		\begin{axis}[
		xlabel={\% Labeled nodes from $\lkg$},
		ylabel={\textbf{Micro-F1}},
		title={Karate Club},
		]
		\plotECNE file[x index=0, y index=0.2] {evaluation/karate/karate_ecne_micro.txt};
			\plotECNED file[x index=0, y index=0.2] {evaluation/karate/karate_ecned_micro.txt};
		\plotNetMF file[x index=0, y index=0.2] {evaluation/karate/karate_NetMF_micro.txt};
		\plotNodeToVec file[x index=0, y index=0.2] {evaluation/karate/karate_node2vec_micro.txt};
		\plotDeepWalk file[x index=0, y index=0.2] {evaluation/karate/karate_deepWalk_micro.txt};
		\plotstructtovec file[x index=0, y index=0.2] {evaluation/karate/karate_struct2vec_micro.txt};
			\plotSDNE file[x index=0, y index=0.2] {evaluation/karate/karate_sdne_micro.txt};
				\plotGraphSage file[x index=0, y index=0.2] {evaluation/karate/karate_graphsage_micro.txt};
		\end{axis}
		\end{tikzpicture}
		\begin{tikzpicture}[scale=\scale][font=\Huge]
		\begin{axis}[
		xlabel={\% Labeled nodes from $\lkg$},
		ylabel={\textbf{Macro-F1}},
		title={Karate Club}
		]
		\plotECNE file[x index=0, y index=0.2] {evaluation/karate/karate_ecne_macro.txt};
				\plotECNED file[x index=0, y index=0.2] {evaluation/karate/karate_ecned_macro.txt};
		\plotNetMF file[x index=0, y index=0.2] {evaluation/karate/karate_NetMF_macro.txt};
		\plotNodeToVec file[x index=0, y index=0.2] {evaluation/karate/karate_node2vec_macro.txt};
		\plotDeepWalk file[x index=0, y index=0.2] {evaluation/karate/karate_deepWalk_macro.txt};
		\plotstructtovec file[x index=0, y index=0.2] {evaluation/karate/karate_struct2vec_macro.txt};
		\plotSDNE file[x index=0, y index=0.2] {evaluation/karate/karate_sdne_macro.txt};
		\plotGraphSage file[x index=0, y index=0.2] {evaluation/karate/karate_graphsage_macro.txt};
		\end{axis}
		\end{tikzpicture}
	\end{minipage}
	\hspace{.5232cm}
	\begin{minipage}{\spaceMiniPage\textwidth}
		\begin{tikzpicture}[scale=\scale][font=\Huge]
		\begin{axis}[
		xlabel={\% Labeled nodes from $\lkg$},
		title={USA Power Grid},
		legend style={at={(-0.6,-0.3)},anchor=west}, 
		]
		\plotECNE file[x index=0, y index=0.2] {evaluation/power/power_ecne_micro.txt};
			\plotECNED file[x index=0, y index=0.2] {evaluation/power/power_ecned_micro.txt};
		\plotNetMF file[x index=0, y index=0.2] {evaluation/power/power_NetMF_micro.txt};
		\plotNodeToVec file[x index=0, y index=0.2] {evaluation/power/power_node2vec_micro.txt};
		\plotDeepWalk file[x index=0, y index=0.2] {evaluation/power/power_deepWalk_micro.txt};
		\plotstructtovec file[x index=0, y index=0.2] {evaluation/power/power_struct2vec_micro.txt};
		\plotSDNE file[x index=0, y index=0.2] {evaluation/power/power_sdne_micro.txt};
		\plotGraphSage file[x index=0, y index=0.2] {evaluation/power/power_graphsage_micro.txt};
		\end{axis}
		\end{tikzpicture}
		\begin{tikzpicture}[scale=\scale][font=\Huge]
		\begin{axis}[
		xlabel={\% Labeled nodes from $\lkg$},
		title={USA Power Grid}
		]
		\plotECNE file[x index=0, y index=0.2] {evaluation/power/power_ecne_macro.txt};
				\plotECNED file[x index=0, y index=0.2] {evaluation/power/power_ecned_macro.txt};
		\plotNetMF file[x index=0, y index=0.2] {evaluation/power/power_NetMF_macro.txt};
		\plotNodeToVec file[x index=0, y index=0.2] {evaluation/power/power_node2vec_macro.txt};
		\plotDeepWalk file[x index=0, y index=0.2] {evaluation/power/power_deepWalk_macro.txt};
		\plotstructtovec file[x index=0, y index=0.2] {evaluation/power/power_struct2vec_macro.txt};
		\plotSDNE file[x index=0, y index=0.2] {evaluation/power/power_sdne_macro.txt};
		\plotGraphSage file[x index=0, y index=0.2] {evaluation/power/power_graphsage_macro.txt};
		\end{axis}
		\end{tikzpicture}
	\end{minipage}
		\hspace{.5232cm}
	\begin{minipage}{\spaceMiniPage\textwidth}
		\begin{tikzpicture}[scale=\scale][font=\Huge]
		\begin{axis}[
		xlabel={\% Labeled nodes from $\lkg$},
		title={Facebook}
		]
		\plotECNE file[x index=0, y index=0.2] {evaluation/facebook/facebook_ecne_micro.txt};
				\plotECNED file[x index=0, y index=0.2] {evaluation/facebook/facebook_ecned_micro.txt};
		\plotNetMF file[x index=0, y index=0.2] {evaluation/facebook/facebook_NetMF_micro.txt};
		\plotNodeToVec file[x index=0, y index=0.2] {evaluation/facebook/facebook_node2vec_micro.txt};
		\plotDeepWalk file[x index=0, y index=0.2] {evaluation/facebook/facebook_deepWalk_micro.txt};
			\plotstructtovec file[x index=0, y index=0.2] {evaluation/facebook/facebook_struct2vec_micro.txt};
		\plotSDNE file[x index=0, y index=0.2] {evaluation/facebook/facebook_sdne_micro.txt};
		\plotGraphSage file[x index=0, y index=0.2] {evaluation/facebook/facebook_graphsage_micro.txt};
		\end{axis}
		\end{tikzpicture}
		\begin{tikzpicture}[scale=\scale][font=\Huge]
		\begin{axis}[
		legend style={at={(-0.3,-0.3)},anchor=west},
		legend columns=-1,
		xlabel={\% Labeled nodes from $\lkg$},
		title={Facebook}
		]
		\plotECNE file[x index=0, y index=0.2] {evaluation/facebook/facebook_ecne_macro.txt};
				\plotECNED file[x index=0, y index=0.2] {evaluation/facebook/facebook_ecned_macro.txt};
		\plotNetMF file[x index=0, y index=0.2] {evaluation/facebook/facebook_NetMF_macro.txt};
		\plotNodeToVec file[x index=0, y index=0.2] {evaluation/facebook/facebook_node2vec_macro.txt};
		\plotDeepWalk file[x index=0, y index=0.2] {evaluation/facebook/facebook_deepWalk_macro.txt};
\plotstructtovec file[x index=0, y index=0.2] {evaluation/facebook/facebook_struct2vec_macro.txt};
\plotSDNE file[x index=0, y index=0.2] {evaluation/facebook/facebook_sdne_macro.txt};
\plotGraphSage file[x index=0, y index=0.2] {evaluation/facebook/facebook_graphsage_macro.txt};
		\end{axis}
		\end{tikzpicture}
	\end{minipage}
		\hspace{.5232cm}
	\begin{minipage}{\spaceMiniPage\textwidth}
		\begin{tikzpicture}[scale=\scale][font=\Huge]
		\begin{axis}[
		xlabel={\% Labeled nodes from $\lkg$},
	ylabel={\textbf{Micro-F1}},
		title={Erdos}
		]
		\plotECNE file[x index=0, y index=0.2] {evaluation/erdos/erdos_ecne_micro.txt};
				\plotECNED file[x index=0, y index=0.2] {evaluation/erdos/erdos_ecned_micro.txt};
		\plotNetMF file[x index=0, y index=0.2] {evaluation/erdos/erdos_NetMF_micro.txt};
		\plotNodeToVec file[x index=0, y index=0.2] {evaluation/erdos/erdos_node2vec_micro.txt};
		\plotDeepWalk file[x index=0, y index=0.2] {evaluation/erdos/erdos_deepWalk_micro.txt};
		\plotstructtovec file[x index=0, y index=0.2] {evaluation/erdos/erdos_struct2vec_micro.txt};
		\plotSDNE file[x index=0, y index=0.2] {evaluation/erdos/erdos_sdne_micro.txt};
		\plotGraphSage file[x index=0, y index=0.2] {evaluation/erdos/erdos_graphsage_micro.txt};
		\end{axis}
		\end{tikzpicture}
		\begin{tikzpicture}[scale=\scale][font=\Huge]
		\begin{axis}[
		legend style={at={(-0.3,-0.3)},anchor=west},
		legend columns=-1,
		xlabel={\% Labeled nodes from $\lkg$},
		ylabel={\textbf{Macro-F1}},
		title={Erdos}
		]
		\plotECNE file[x index=0, y index=0.2] {evaluation/erdos/erdos_ecne_macro.txt};
				\plotECNED file[x index=0, y index=0.2] {evaluation/erdos/erdos_ecned_macro.txt};
		\plotNetMF file[x index=0, y index=0.2] {evaluation/erdos/erdos_NetMF_macro.txt};
		\plotNodeToVec file[x index=0, y index=0.2] {evaluation/erdos/erdos_node2vec_macro.txt};
		\plotDeepWalk file[x index=0, y index=0.2] {evaluation/erdos/erdos_deepWalk_macro.txt};
			\plotstructtovec file[x index=0, y index=0.2] {evaluation/erdos/erdos_struct2vec_macro.txt};
		\plotSDNE file[x index=0, y index=0.2] {evaluation/erdos/erdos_sdne_macro.txt};
		\plotGraphSage file[x index=0, y index=0.2] {evaluation/erdos/erdos_graphsage_macro.txt};
		\end{axis}
		\end{tikzpicture}
	\end{minipage}
			\hspace{.5232cm}
	\begin{minipage}{\spaceMiniPage\textwidth}
		\begin{tikzpicture}[scale=\scale][font=\Huge]
		\begin{axis}[
		xlabel={\% Labeled nodes from $\lkg$},
		title={Astrophysics}
		]
		\plotECNE file[x index=0, y index=0.2] {evaluation/astro/astro_ecne_micro.txt};
				\plotECNED file[x index=0, y index=0.2] {evaluation/astro/astro_ecned_micro.txt};
		\plotNetMF file[x index=0, y index=0.2] {evaluation/astro/astro_NetMF_micro.txt};
		\plotNodeToVec file[x index=0, y index=0.2] {evaluation/astro/astro_node2vec_micro.txt};
		\plotDeepWalk file[x index=0, y index=0.2] {evaluation/astro/astro_deepWalk_micro.txt};
			\plotstructtovec file[x index=0, y index=0.2] {evaluation/astro/astro_struct2vec_micro.txt};
		\plotSDNE file[x index=0, y index=0.2] {evaluation/astro/astro_sdne_micro.txt};
		\plotGraphSage file[x index=0, y index=0.2] {evaluation/astro/astro_graphsage_micro.txt};
		\end{axis}
		\end{tikzpicture}
		\begin{tikzpicture}[scale=\scale][font=\Huge]
		\begin{axis}[
		legend style={at={(-0.3,-0.3)},anchor=west},
		legend columns=-1,
		xlabel={\% Labeled nodes from $\lkg$},
		title={Astrophysics}
		]
		\plotECNE file[x index=0, y index=0.2] {evaluation/astro/astro_ecne_macro.txt};
			\plotECNED file[x index=0, y index=0.2] {evaluation/astro/astro_ecned_macro.txt};
		\plotNetMF file[x index=0, y index=0.2] {evaluation/astro/astro_NetMF_macro.txt};
		\plotNodeToVec file[x index=0, y index=0.2] {evaluation/astro/astro_node2vec_macro.txt};
		\plotDeepWalk file[x index=0, y index=0.2] {evaluation/astro/astro_deepWalk_macro.txt};
			\plotstructtovec file[x index=0, y index=0.2] {evaluation/astro/astro_struct2vec_macro.txt};
		\plotSDNE file[x index=0, y index=0.2] {evaluation/astro/astro_sdne_macro.txt};
		\plotGraphSage file[x index=0, y index=0.2] {evaluation/astro/astro_graphsage_macro.txt};
		\end{axis}
		\end{tikzpicture}
	\end{minipage}
			\hspace{.5232cm}
	\begin{minipage}{\spaceMiniPage\textwidth}
		\begin{tikzpicture}[scale=\scale][font=\Huge]
		\begin{axis}[
		xlabel={\% Labeled nodes from $\lkg$},
		title={ArXiv}
		]
		\plotECNE file[x index=0, y index=0.2] {evaluation/arxiv/arxiv_ecne_micro.txt};
				\plotECNED file[x index=0, y index=0.2] {evaluation/arxiv/arxiv_ecned_micro.txt};
		\plotNetMF file[x index=0, y index=0.2] {evaluation/arxiv/arxiv_NetMF_micro.txt};
		\plotNodeToVec file[x index=0, y index=0.2] {evaluation/arxiv/arxiv_node2vec_micro.txt};
		\plotDeepWalk file[x index=0, y index=0.2] {evaluation/arxiv/arxiv_deepWalk_micro.txt};
			\plotstructtovec file[x index=0, y index=0.2] {evaluation/arxiv/arxiv_struct2vec_micro.txt};
		\plotSDNE file[x index=0, y index=0.2] {evaluation/arxiv/arxiv_sdne_micro.txt};
		\plotGraphSage file[x index=0, y index=0.2] {evaluation/arxiv/arxiv_graphsage_micro.txt};
		\end{axis}
		\end{tikzpicture}
		\begin{tikzpicture}[scale=\scale][font=\Huge]
		\begin{axis}[
		legend style={at={(-0.3,-0.3)},anchor=west},
		legend columns=-1,
		xlabel={\% Labeled nodes from $\lkg$},
		title={ArXiv}
		]
		\plotECNE file[x index=0, y index=0.2] {evaluation/arxiv/arxiv_ecne_macro.txt};
				\plotECNED file[x index=0, y index=0.2] {evaluation/arxiv/arxiv_ecned_macro.txt};
		\plotNetMF file[x index=0, y index=0.2] {evaluation/arxiv/arxiv_NetMF_macro.txt};
		\plotNodeToVec file[x index=0, y index=0.2] {evaluation/arxiv/arxiv_node2vec_macro.txt};
		\plotDeepWalk file[x index=0, y index=0.2] {evaluation/arxiv/arxiv_deepWalk_macro.txt};
			\plotstructtovec file[x index=0, y index=0.2] {evaluation/arxiv/arxiv_struct2vec_macro.txt};
		\plotSDNE file[x index=0, y index=0.2] {evaluation/arxiv/arxiv_sdne_macro.txt};
		\plotGraphSage file[x index=0, y index=0.2] {evaluation/arxiv/arxiv_graphsage_macro.txt};
		
		\end{axis}
		\end{tikzpicture}
	\end{minipage}
	\caption{Edge classification on varying the amount of labeled data used for training (average of 5 runs).}
	\label{fig:classification-homo}
\end{figure*}
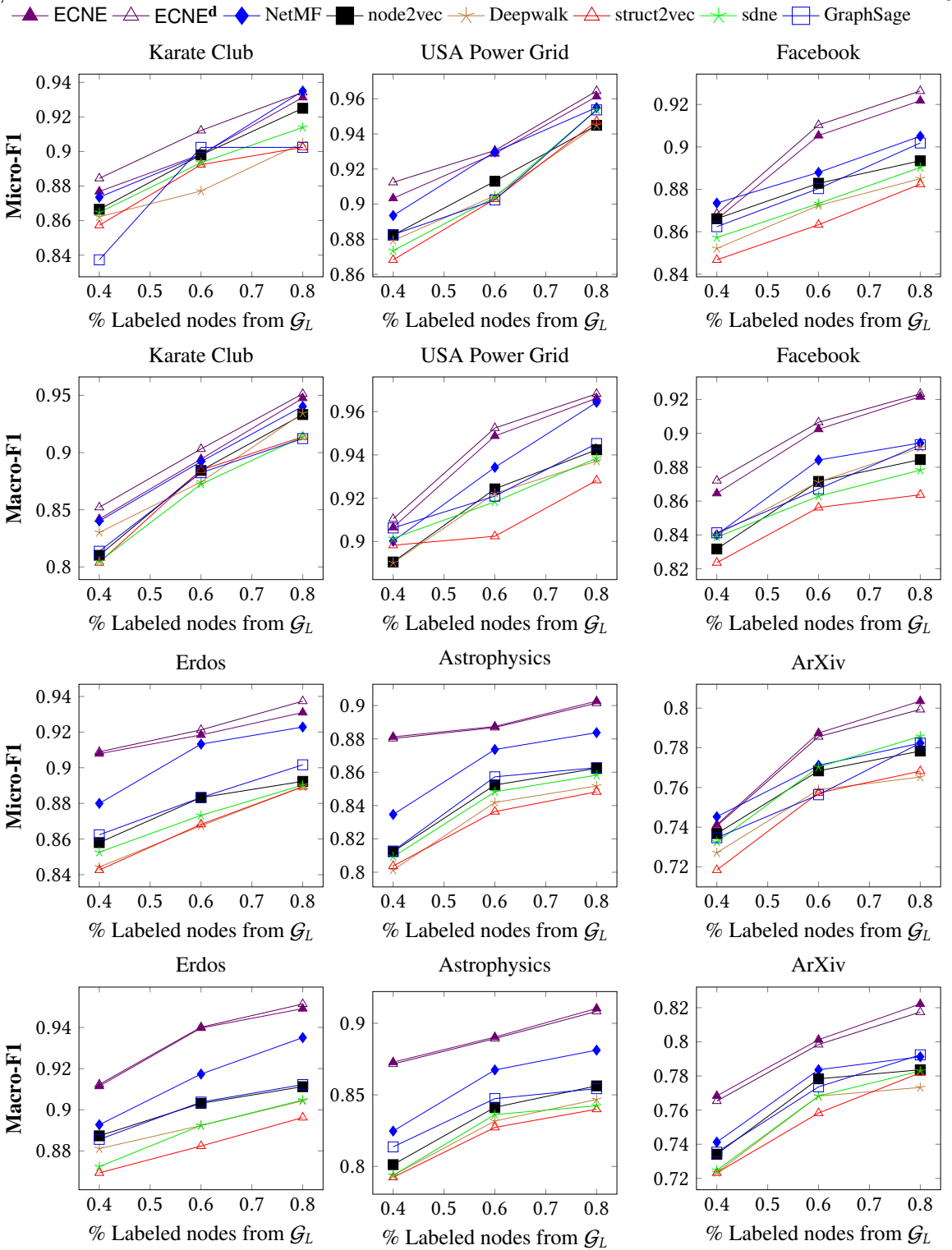

\begin{figure*}[!h]
	\centering
	\begin{tikzpicture}
	\begin{customlegend}[legend columns=8,legend style={align=left,draw=none,column sep=0ex},legend entries={\system, \systemD, NetMF, node2vec, Deepwalk, struct2vec, sdne, GraphSage}]
	\addlegendimage{color=violet!90!black, fill=violet!90!black,mark=square*,  only marks, mark size=4pt}
	\addlegendimage{color=pink, fill=pink,mark=square*,  only marks, mark size=4pt}
	\addlegendimage{color=blue, fill=blue,mark=square*,  only marks, mark size=4pt}
	\addlegendimage{color=black, fill=black,mark=square*,  only marks, mark size=4pt}
	\addlegendimage{color=brown, fill=brown,mark=square*,  only marks, mark size=4pt}
	\addlegendimage{color=red, fill=red,mark=square*,  only marks, mark size=4pt}
	\addlegendimage{color=green!90!black, fill=green!90!black,mark=square*,  only marks, mark size=4pt}
	\addlegendimage{color=yellow, fill=yellow,mark=square*,  only marks, mark size=4pt}
	\end{customlegend}
	\end{tikzpicture}
	\begin{minipage}{\textwidth}
		\begin{tikzpicture}[scale=\scaleClust][font=\Large]
		\begin{axis}[
		width  = 0.5*\textwidth,
		height = 4.2cm,
		xlabel = {Karate Club},
		ybar=2*\pgflinewidth,
		bar width=10pt,
		ymajorgrids = true,
		symbolic x coords={T,T1,N,D,E,S,ST,GS},
		scaled y ticks = false,
		enlarge x limits=1,
		ymin=0,
		xmajorticks=false,
		legend pos=outer north east]
		]
		\plotECNEH     coordinates {(T,1.0) +- (0.0, 0.5)};
		\plotECNEDH     coordinates {(T1,1.0)};
		\plotNetMFH coordinates {(N,0.98834)};
		\plotNodeToVecH coordinates {(D,0.89556)};
		\plotDeepWalkH coordinates {(E,0.895356)};
		\plotSDNEH coordinates {(S,0.901373)};
		\plotstructtovecH coordinates {(ST,0.89214)};
		\plotGraphSageH coordinates {(GS,0.93456)};
		\end{axis}
		\end{tikzpicture}
		\begin{tikzpicture}[scale=\scaleClust][font=\Large]
		\begin{axis}[
		width  = 0.5*\textwidth,
		height = 4.2cm,
		xlabel = {Erdos},
		ybar=2*\pgflinewidth,
		bar width=10pt,
		ymajorgrids = true,
		symbolic x coords={T,T1,N,D,E,S,ST,GS}, 
		scaled y ticks = false,
		enlarge x limits=1,
		ymin=0,
		xmajorticks=false,
		legend pos=outer north east]
		]
		\plotECNEH     coordinates {(T,.955)};
		\plotECNEDH   coordinates {(T1,.9612)};
		\plotNetMFH coordinates {(N,0.93234)};
		\plotNodeToVecH coordinates {(D,0.9056)};
		\plotDeepWalkH coordinates {(E,0.89374)};
		\plotSDNEH coordinates {(S,0.901234)};
		\plotstructtovecH coordinates {(ST,0.912323)};
		\plotGraphSageH coordinates {(GS,0.92151)};
		\end{axis}
		\end{tikzpicture}
	\end{minipage}
	\begin{minipage}{\textwidth}
		\begin{tikzpicture}[scale=\scaleClust][font=\Large]
		\begin{axis}[
		width  = 0.5*\textwidth,
		height = 4.2cm,
		xlabel = {Facebook},
		ybar=2*\pgflinewidth,
		bar width=10pt,
		ymajorgrids = true,
		symbolic x coords={T,T1,N,D,E,S,ST,GS}, 
		scaled y ticks = false,
		enlarge x limits=1,
		ymin=0,
		y tick label style={
			/pgf/number format/fixed
		},
		xmajorticks=false,
		legend cell align=left,
		legend style={
			at={(0.5,-0.2)},
			anchor=north,legend columns=-1,
			column sep=1ex
		}
		]
		\plotECNEH     coordinates {(T,0.95)};
		\plotECNEDH    coordinates {(T1,.937)};
		\plotNetMFH coordinates {(N,0.8934)};
		\plotNodeToVecH coordinates {(D,0.8836)};
		\plotDeepWalkH coordinates {(E,0.88126)};
		\plotSDNEH coordinates {(S,0.934)};
		\plotstructtovecH coordinates {(ST,0.8734)};
		\plotGraphSageH coordinates {(GS,0.8421)};
		\end{axis}
		\end{tikzpicture}
		\begin{tikzpicture}[scale=\scaleClust][font=\Large]
		\begin{axis}[
		width  = 0.5*\textwidth,
		height = 4.2cm,
		xlabel = {Power Grid},
		ybar=2*\pgflinewidth,
		bar width=10pt,
		ymajorgrids = true,
		symbolic x coords={T,T1,N,D,E,S,ST,GS}, 
		scaled y ticks = false,
		enlarge x limits=1,
		ymin=0,
		xmajorticks=false,
		legend pos=outer north east]
		]
		\plotECNEH     coordinates {(T,.932)};
		\plotECNEDH    coordinates {(T1,.9427)};
		\plotNetMFH coordinates {(N,0.934)};
		\plotNodeToVecH coordinates {(D,0.8734)};
		\plotDeepWalkH coordinates {(E,0.8421)};
		\plotSDNEH coordinates {(S,0.88325)};
		\plotstructtovecH coordinates {(ST,0.897334)};
		\plotGraphSageH coordinates {(GS,0.9013)};
		\end{axis}
		\end{tikzpicture} 
	\end{minipage}
	\begin{minipage}{\textwidth}
		\begin{tikzpicture}[scale=\scaleClust][font=\Large]
		\begin{axis}[
		width  = 0.5*\textwidth,
		height = 4.2cm,
		xlabel = {Astrophysics},
		ybar=2*\pgflinewidth,
		bar width=10pt,
		ymajorgrids = true,
		symbolic x coords={T,T1,N,D,E,S,ST,GS}, 
		scaled y ticks = false,
		enlarge x limits=1,
		ymin=0,
		xmajorticks=false,
		legend pos=outer north east]
		]
		\plotECNEH     coordinates {(T,0.86463)};
		\plotECNEDH    coordinates {(T1,.85621)};
		\plotNetMFH coordinates {(N,0.8218)};
		\plotNodeToVecH coordinates {(D,0.8012)};
		\plotDeepWalkH coordinates {(E,0.7936)};
		\plotSDNEH coordinates {(S,0.8341)};
		\plotstructtovecH coordinates {(ST,0.84134)};
		\plotGraphSageH coordinates {(GS,0.8618)};
		\end{axis}
		\end{tikzpicture}
		\begin{tikzpicture}[scale=\scaleClust][font=\Large]
		\begin{axis}[
		width  = 0.5*\textwidth,
		height = 4.2cm,
		xlabel = {ArXiv},
		ybar=2*\pgflinewidth,
		bar width=10pt,
		ymajorgrids = true,
		symbolic x coords={T,T1,N,D,E,S,ST,GS}, 
		scaled y ticks = false,
		enlarge x limits=1,
		ymin=0,
		xmajorticks=false,
		legend pos=outer north east]
		]
		\plotECNEH     coordinates {(T,0.784545)};
		\plotECNEDH    coordinates {(T1,.7781)};
		\plotNetMFH coordinates {(N,0.745253)};
		\plotNodeToVecH coordinates {(D,0.732424)};
		\plotDeepWalkH coordinates {(E,0.72343)};
		\plotSDNEH coordinates {(S,0.76564)};
		\plotstructtovecH coordinates {(ST,0.77212)};
		\plotGraphSageH coordinates {(GS,0.78216)};
		\end{axis}
		\end{tikzpicture}
	\end{minipage}
	\caption{Clustering results (average of 5 runs).}
	\label{fig:clustering}
\end{figure*}
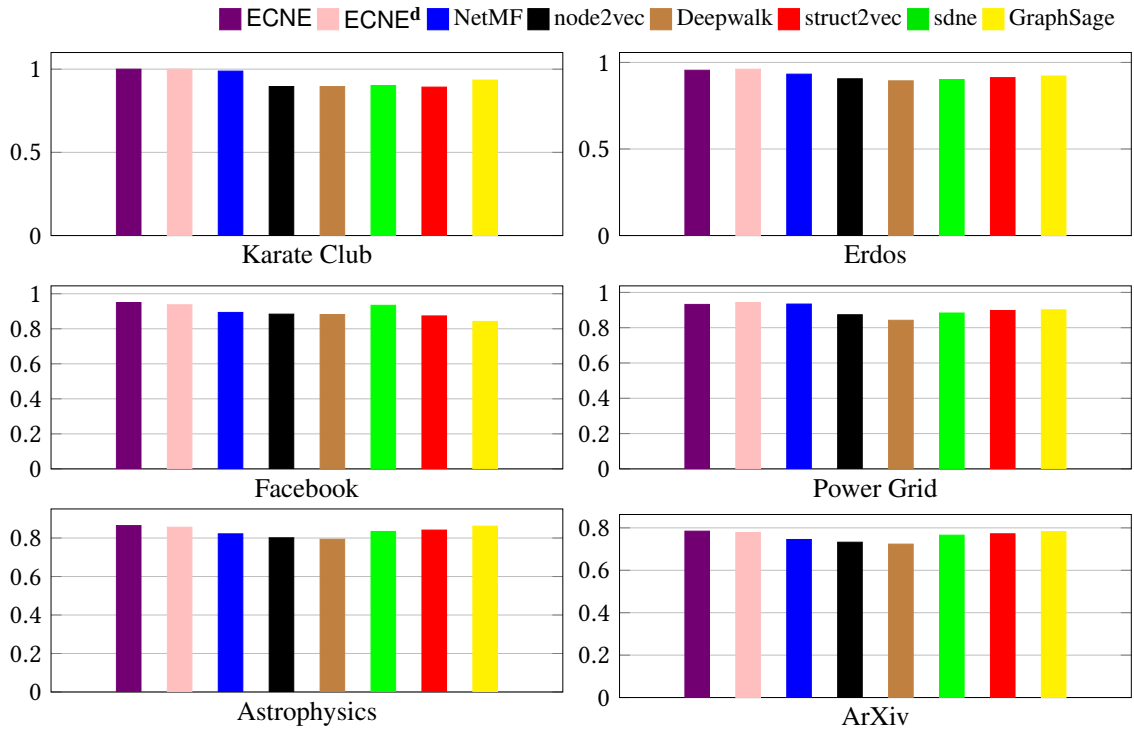
We are not aware of any approach that focuses on directly learning edge embeddings. Therefore, we adapt a sample of the most popular and well-performing node embedding systems to construct edge embeddings as done by Grover and Leskovec \cite{grover2016node2vec}. Given an edge $e=(u,v)$, its embedding is obtained as \embE{$e$}=\emb{$u$} $\circ$\emb{$v$}, where $\circ$ is the average, Hadamard product, Weighted-L or Weighted-L2 operator. We considered the following competitors:
\begin{itemize}
	\item \textbf{DeepWalk} learns node embeddings via random walks  and the Skip-gram model. 
	\item \textbf{node2vec} improves upon DeepWalk in both the way random walks are generated (by balancing the breadth-first search and depth-first search strategies).
	\item \textbf{NetMF}~\cite{qiu2018network} is a very recent piece of work, which shows that Deepwalk, LINE, and node2vec with negative sampling can be unified into the matrix factorization framework with closed forms\footnote{https://github.com/xptree/NetMF}.
	\item \textbf{struct2vec}~\cite{ribeiro2017struc2vec} is a recent approach also based on random walks, which imposes similar embeddings to nodes that are structurally similar\footnote{https://github.com/leoribeiro/struc2vec}. 
	\item \textbf{SDNE}~\cite{wang2016structural} uses autoencoders to preserve structural similarity\footnote{https://github.com/suanrong/SDNE}.
	\item \textbf{GraphSage}~\cite{hamilton2017inductive} aggregates fixed-size neighborhood information of nodes\footnote{https://github.com/williamleif/GraphSAGE}.
\end{itemize}

We used the values of the parameters $n$ (number of walks per node), $L$ (max. walk length), $w$ (window size for the Skipgram model), and negative samples ($\Gamma$) in line with values used by the competitors as reported in their respective paper. Ditto for specific parameters  (e.g., $p$ and $q$ for node2vec). We set $n$=$10$, $L$=100 $w$=10, and $\Gamma$=100 for all systems in all experiments.

\subsection{Setting the embedding dimension} 
A different strategy is adopted for the dimension of the embeddings $d$. On one hand, we note that the number of node embeddings found by the competitors corresponds to the number of nodes of the networks. In this case, we used  $d$=128 for all datasets, which is in line with values used by recent competitors (e.g., NetMF \cite{qiu2018network}). 

On the other hand, we observe that for \system\ the number of nodes (of the weighted line graph) to embed corresponds to the number of edges of the original graph, which is usually larger. This inherently makes the number of parameters for \system\ larger. Therefore, for \system, we performed experiments in two different settings. The first assumes $d$=128 in all datasets,
 while in the second we adapted the embedding dimensionality for each dataset in order for \system\ to use the same number of parameters as the competitors. 
 
 We refer to this variant of \system\ as \systemD.  For instance, for the Erdos dataset including 6100 nodes, the number of parameters used by the competitors was 6100*128=780800. In order for \system\ to use (roughly) the same number of parameters we have to consider $d=80$ as the number of edges 9939 multiplied by 80 gives 795120.
\subsection{Community Detection via Edge Embeddings}
\label{sec:exp-edge-embeddings}
We start by presenting experiments in the task of community detection via edge clustering. While Evans and Lambiotte \cite{evans2009line} treated the problem from a modularity-optimization perspective, our goal is to tackle this problem from an edge-embedding perspective and show its usefulness. To conduct experiments, we proceeded as follows. For each considered network in Table \ref{tab:datasets}, we first found communities by using a modularity-based algorithm \cite{newman2006modularity}. Then, for each community, intended as a set of nodes, we identified the set of intra-community edges and labeled each of such edges with the {id} of the community it belongs to. To evaluate \system\ and competitors we computed the edge embeddings for each network; while for \system\ (and \systemD) edge embeddings are directly computed, for the other approaches we used the bootstrapping approach previously described (\cref{sec:exp-setting}). After obtaining edge embeddings for all approaches, we trained a one-vs-rest Logistic regression model giving as input the edge embeddings and the labels (the community they belong to) and computed the Micro and Macro F1 scores. 

Results when varying the amount of labeled data used for training are reported in Fig. \ref{fig:classification-homo}. We point out that for the competitors the figure reports the best edge embeddings obtained over all the operators on node embeddings considered (\cref{sec:exp-setting}). In particular, we observed that for node2vec and Deepwalk in most of the cases (e.g., on Astrophysics and ArXiv) the average of the node embeddings gave the best results while for NetFM, struct2vec, SDNE, and GraphSage in some cases (e.g., on ArXiv) the Weighted-L2 performed better.
Fig. \ref{fig:classification-homo} shows that \systemD\ performs better than the competitors in almost all cases. We recall that this variant of \system, instead of considering a fixed embedding size (i.e., $d$=128) adapts the dimension to the number of edges in the network (\cref{sec:exp-setting}). It seems that for networks of moderate size ($\sim$10K edges), the size of embeddings equal to 128 leads to slightly inferior results. However, when the size of the network (in terms of edges) increases (i.e., on Astrophysics and ArXiv) we note that \system\ performs slightly better. This may be explained by the fact that the valu of $d$ automatically set may not be enough to correctly separate edge embeddings. As an example, to obtain the same number of parameters as the competitors on ArXiv, \systemD\ used $d$=20 instead of $d$=128. We also observe that when moving to larger networks the difference wrt the competitors of both \system\ variants becomes clearer.

\subsection{Experiments on edge clustering}
We also completed experiments in an edge clustering task. We considered the K-means algorithm to which we gave as input both the edge embeddings obtained by \system\ (in its two variant) and competitors and the number of clusters. In particular, the number of clusters considered coincided with the number of communities previously found. To evaluate the performance of the systems, we compute the Normalized Mutual Information (NMI), which is used to estimate the clustering quality. Results are reported in Fig. \ref{fig:clustering}. We observe that \system\ and \systemD\ perform equally or better than the competitors in all networks. 

We observe that also in this case the improvement wrt the competitors becomes clearer as the network size increases. Note that we only report the best performance for the competitors in terms of the aggregation mechanism on node embeddings that, even in this case, in most cases was the average. On the contrary, neither \system\ nor \systemD\ require aggregation as they directly learns edge embeddings.
As observed in the previous experiments, even in this case we note that \systemD\ performs slightly better than \system\ for moderately large networks. As the number of  edges of the network increases, the difference in performance is no more tangible.

We mention that the main goal of this paper neither is to specifically devise a community detection algorithm nor a clustering one. The goal is to introduce the novel task of edge-centric network embeddings and show its usefulness in concrete downstream applications. This leaves room for further investigations related to edge-driven community detection and clustering. Examples are alternative weighing mechanisms for the line graph or a different way of computing embeddings from the line graph (e.g., using SDNE instead of the Skip-graph model). 
\begin{table*}[!h]
	\centering
	\begin{tabular}{@{}cc @{\hspace{.001\tabcolsep}} c@{\hspace{.001\tabcolsep}}c@{\hspace{.001\tabcolsep}}c@{\hspace{.001\tabcolsep}}
			c@{\hspace{.001\tabcolsep}}c@{\hspace{.001\tabcolsep}}c@{\hspace{.001\tabcolsep}}
			c@{}}
		\toprule
		\textbf{Approach} \\(st. dev.)                  & \textbf{USAir}   & 
		\textbf{NS}    & \textbf{PB}     & 
		\textbf{CEl}  & 
		\textbf{EColi} &     \textbf{ArX} & \textbf{DBLP} &\multirow{8}{*}{}  \\
		\cmidrule(r){1-8}  \systemLP-LSTM &    \multicolumn{1}{c}{{.964}} 
		& 
		\multicolumn{1}{c} {.981} & \multicolumn{1}{c}{{.831}} & 
		\multicolumn{1}{c}{{.962}} & 
		\multicolumn{1}{c}{{.924}} &   \multicolumn{1}{c}{{\textbf{.995}}}   &.978              \\ 
		&    \multicolumn{1}{c}{.014} 
		& 
		\multicolumn{1}{c} {.011} & \multicolumn{1}{c}{{.088}} & 
		\multicolumn{1}{c}{{.033}} & 
		\multicolumn{1}{c}{{.071}} &   \multicolumn{1}{c}{{.003}}         & \multicolumn{1}{c}{{.013}}          \\ 
		\cmidrule(r){1-8}
		\cmidrule(r){1-8}\systemLP-Max &    \multicolumn{1}{c}{.921} 
		& 
		\multicolumn{1}{c} {.967} & \multicolumn{1}{c}{.803} & 
		\multicolumn{1}{c}{.954} & 
		\multicolumn{1}{c}{.912} &         \multicolumn{1}{c}{.967}  &.\textbf{979 }         \\ 
		&    \multicolumn{1}{c}{.062} 
		& 
		\multicolumn{1}{c} {.028} & \multicolumn{1}{c}{{.039}} & 
		\multicolumn{1}{c}{{.075}} & 
		\multicolumn{1}{c}{{.071}} &   \multicolumn{1}{c}{{.021}}      &.014           \\ 
		\cmidrule(r){1-8} 
		\cmidrule(r){1-8}\systemLP-Avg &    \multicolumn{1}{c}{.912} 
		& 
		\multicolumn{1}{c} {.965} & \multicolumn{1}{c}{.812} & 
		\multicolumn{1}{c}{.957} & 
		\multicolumn{1}{c}{.911} &   \multicolumn{1}{c}{.957}      &.948           \\ 
		&    \multicolumn{1}{c}{.035} 
		& 
		\multicolumn{1}{c} {.019} & \multicolumn{1}{c}{{.071}} & 
		\multicolumn{1}{c}{{.027}} & 
		\multicolumn{1}{c}{{.080}} &   \multicolumn{1}{c}{{.042}}     &.021            \\ 
		\cmidrule(r){1-8}  \systemD-LP-LSTM &    \multicolumn{1}{c}{\textbf{.971}} 
		& 
		\multicolumn{1}{c} {\textbf{.983}} & \multicolumn{1}{c}{{.836}} & 
		\multicolumn{1}{c}{{\textbf{.967}}} & 
		\multicolumn{1}{c}{{.928}} &   \multicolumn{1}{c}{{.993}}     &.976            \\ 
		&    \multicolumn{1}{c}{.012} 
		& 
		\multicolumn{1}{c} {.016} & \multicolumn{1}{c}{{.068}} & 
		\multicolumn{1}{c}{{.022}} & 
		\multicolumn{1}{c}{{.066}} &   \multicolumn{1}{c}{{.002}}        &.022         \\ 
		\cmidrule(r){1-8}\systemD-LP-Max &    \multicolumn{1}{c}{.922} 
		& 
		\multicolumn{1}{c} {.967} & \multicolumn{1}{c}{.804} & 
		\multicolumn{1}{c}{.956} & 
		\multicolumn{1}{c}{.913} &         \multicolumn{1}{c}{.968}      &.965      \\ 
		&    \multicolumn{1}{c}{.063} 
		& 
		\multicolumn{1}{c} {.026} & \multicolumn{1}{c}{{.069}} & 
		\multicolumn{1}{c}{{.023}} & 
		\multicolumn{1}{c}{{.027}} &   \multicolumn{1}{c}{{.021}}      &.032           \\ 
		\cmidrule(r){1-8}\systemD-LP-Avg &    \multicolumn{1}{c}{.913} 
		& 
		\multicolumn{1}{c} {.966} & \multicolumn{1}{c}{\textbf{.838}} & 
		\multicolumn{1}{c}{.959} & 
		\multicolumn{1}{c}{.912} &   \multicolumn{1}{c}{.960}          &.951       \\ 
		&    \multicolumn{1}{c}{.014} 
		& 
		\multicolumn{1}{c} {.028} & \multicolumn{1}{c}{{.068}} & 
		\multicolumn{1}{c}{{.022}} & 
		\multicolumn{1}{c}{{.076}} &   \multicolumn{1}{c}{{.028}}      &.031           \\ 
		\cmidrule(r){1-8} LEAP-LSTM&    \multicolumn{1}{c}{.962} 
		& 
		\multicolumn{1}{c} {{.982}} & \multicolumn{1}{c}{.814} & 
		\multicolumn{1}{c}{.957} & 
		\multicolumn{1}{c}{\textbf{.926}} &      \multicolumn{1}{c}{.994}  &.976              \\ 
		&    \multicolumn{1}{c}{.024} 
		& 
		\multicolumn{1}{c} {.011} & \multicolumn{1}{c}{{.072}} & 
		\multicolumn{1}{c}{{.023}} & 
		\multicolumn{1}{c}{{.063}} &   \multicolumn{1}{c}{{.001}}     &.002            \\ 
		
		\cmidrule(r){1-8} LEAP-Max&    \multicolumn{1}{c}{.949} 
		& 
		\multicolumn{1}{c} {{.942}} & \multicolumn{1}{c}{.804} & 
		\multicolumn{1}{c}{.927} & 
		\multicolumn{1}{c}{.913} &      \multicolumn{1}{c}{.974}       &.966         \\ 
		&    \multicolumn{1}{c}{.026} 
		& 
		\multicolumn{1}{c} {.035} & \multicolumn{1}{c}{{.069}} & 
		\multicolumn{1}{c}{{.071}} & 
		\multicolumn{1}{c}{{.072}} &   \multicolumn{1}{c}{{.021}}   &.011              \\ 
		\cmidrule(r){1-8} LEAP-Avg&    \multicolumn{1}{c}{.933} 
		& 
		\multicolumn{1}{c} {{.962}} & \multicolumn{1}{c}{.798} & 
		\multicolumn{1}{c}{.947} & 
		\multicolumn{1}{c}{.903} &      \multicolumn{1}{c}{.964}   &.948             \\ 
		&    \multicolumn{1}{c}{.016} 
		& 
		\multicolumn{1}{c} {.071} & \multicolumn{1}{c}{{.078}} & 
		\multicolumn{1}{c}{{.024}} & 
		\multicolumn{1}{c}{{.078}} &   \multicolumn{1}{c}{{.021}}      &.014           \\ 

		\cmidrule(r){1-8}NetMF&    \multicolumn{1}{c}{.887} 
		& 
		\multicolumn{1}{c} {.856} & \multicolumn{1}{c}{.797} & 
		\multicolumn{1}{c}{.826} & 
		\multicolumn{1}{c}{.907} &     \multicolumn{1}{c}{.947}    &.938             \\ 
		&    \multicolumn{1}{c}{.076} 
		& 
		\multicolumn{1}{c} {.102} & \multicolumn{1}{c}{{.128}} & 
		\multicolumn{1}{c}{{.131}} & 
		\multicolumn{1}{c}{{.022}} &   \multicolumn{1}{c}{{.026}}     &.081            \\ 
		%
		\cmidrule(r){1-9}node2vec&    \multicolumn{1}{c}{.865} 
		& 
		\multicolumn{1}{c} {.824} & \multicolumn{1}{c}{.767} & 
		\multicolumn{1}{c}{.811} & 
		\multicolumn{1}{c}{.887} &   \multicolumn{1}{c}{.921}          &.876       \\ 
		&    \multicolumn{1}{c}{.014} 
		& 
		\multicolumn{1}{c} {.111} & \multicolumn{1}{c}{{.182}} & 
		\multicolumn{1}{c}{{.113}} & 
		\multicolumn{1}{c}{{.109}} &   \multicolumn{1}{c}{{.031}}    &.103             \\ 
		%
		\cmidrule(r){1-8}Deepwalk&    \multicolumn{1}{c}{.847} 
		& 
		\multicolumn{1}{c} {.813} & \multicolumn{1}{c}{.772} & 
		\multicolumn{1}{c}{.796} & 
		\multicolumn{1}{c}{.869} &       \multicolumn{1}{c}{.902}       &.931       \\ 
		&    \multicolumn{1}{c}{.068} 
		& 
		\multicolumn{1}{c} {.080} & \multicolumn{1}{c}{{.121}} & 
		\multicolumn{1}{c}{{.141}} & 
		\multicolumn{1}{c}{{.115}} &   \multicolumn{1}{c}{{.063}}     &.012            \\ 
		%
		\cmidrule(r){1-8}struct2vec&    \multicolumn{1}{c}{.818} 
		& 
		\multicolumn{1}{c} {.807} & \multicolumn{1}{c}{.711} & 
		\multicolumn{1}{c}{.921} & 
		\multicolumn{1}{c}{.901} &       \multicolumn{1}{c}{.931}    &.897   \\ 
		&    \multicolumn{1}{c}{.121} 
		& 
		\multicolumn{1}{c} {.132} & \multicolumn{1}{c}{{.021}} & 
		\multicolumn{1}{c}{{.064}} & 
		\multicolumn{1}{c}{{.079}} &   \multicolumn{1}{c}{{.042}}     &.067            \\ 
		\cmidrule(r){1-8}sdne&    \multicolumn{1}{c}{.852} 
		& 
		\multicolumn{1}{c} {.857} & \multicolumn{1}{c}{.802} & 
		\multicolumn{1}{c}{.924} & 
		\multicolumn{1}{c}{.912} &       \multicolumn{1}{c}{.961}    &.911  
		\\ 
		&    \multicolumn{1}{c}{.050} 
		& 
		\multicolumn{1}{c} {.089} & \multicolumn{1}{c}{{.041}} & 
		\multicolumn{1}{c}{{.054}} & 
		\multicolumn{1}{c}{{.084}} &   \multicolumn{1}{c}{{.029}}    &.037             \\ 
		\cmidrule(r){1-8}GraphSage&    \multicolumn{1}{c}{.872} 
		& 
		\multicolumn{1}{c} {.859} & \multicolumn{1}{c}{.793} & 
		\multicolumn{1}{c}{.952} & 
		\multicolumn{1}{c}{.907} &       \multicolumn{1}{c}{.961}   &.911    \\      
		&    \multicolumn{1}{c}{.107} 
		& 
		\multicolumn{1}{c} {.111} & \multicolumn{1}{c}{{.143}} & 
		\multicolumn{1}{c}{{.023}} & 
		\multicolumn{1}{c}{{.087}} &   \multicolumn{1}{c}{{.012}}      &.041           \\ 
		\bottomrule
	\end{tabular}
	\caption{{Area under the ROC curve (AUC) and standard deviation.} }
	\label{tab:results-AUC}
	\vspace{-.2cm}
\end{table*}
\subsection{Link Prediction}
\label{sec:exp-link-prediction}
In this set of experiments, the goal was as follows: given a graph $\kgDef$ and a pair of nodes $(u, v)$, what is the probability of the existence of the link between $u$ and $v$? To set-up the learning model described in Section~\ref{sec:link-prediction}, we can consider each existing edge  $(x, y) \in V_G$ as a positive example while negative examples can be sampled such that the edge $(\overline{x}, \overline{y}) \notin V_G$. With this reasoning, we assign as a label 1 to positive pairs and 0 to negative pairs.

To evaluate our proposal and compare it with competitors, we sampled a variable number of both positive and negative edges and further split them into train and test examples according to the state-of-the-art~\cite{agrawal2019learning}. In particular, for small datasets, we leverage 90\% of edges for train and the remaining for test also considering an equal number of negative edges. For datasets with more than 4K nodes, we adopt a 50\% split. For our approach and LEAP (the other competitor using paths) we considered as path lengths $l=3$ and $l=4$ and used up to 100 paths for each such lengths randomly selected. To train the model, we used the Adam optimizer (and learning rate of 0.001) with binary cross-entropy for a maximum of 50 epochs with early-stop enabled. For the other competitors, we used the best configuration reported in their respective papers~\cite{agrawal2019learning}. 

{The competitors considered are node2vec, Deepwalk, NetFM} that do not make usage of multi-hop information. We also considered SDNE and struct2vec that learn node embeddings to preserve structural similarity and GraphSage, which incorporates neighborhood information via aggregation. However, our direct competitor is the state-of-the-art {LEAP}~\cite{agrawal2019learning} system, which uses embeddings on nodes in a path along with path aggregation but does not consider edge embeddings.  We have also considered variants of \systemLP\ that instead of considering edge embeddings computed via \system\ (or \systemD), compute them as done by the other competitors. In particular, we considered the average of node embeddings and the LSTM-based path aggregator. However, we found that in all cases results were inferior to the case in which embeddings were computed via \system\ or \systemD. Therefore, for sake of space we omit these results. 

\noindent
\textbf{Results.} Table \ref{tab:results-AUC} reports the results. We observe that approaches based on paths that leverage multi-hop information (rows 1-10) perform better on the task of link prediction. This comes as no surprise as the presence/absence of a link can be better understood by looking at the overall connectivity between a pair of candidate nodes for the link. We also note that computing edges with \systemD-LP gives better results for moderately large networks. On the two larger networks, \systemLP, which sets the embedding dimensions to 128 performs better. This is in line with results obtained on edge-based community detection and edge clustering. We also observe that the aggregation strategy considering paths as sequences of edges (all \system\ variants) gives better performance than when considering them as sequences of nodes (all LEAP variants) in almost all the cases. 

\section{Concluding Remarks and Future Work}
\label{sec:conclusions}
We introduced the task of edge-centric network embeddings. The main intuition of our approach is to directly compute edge embeddings by transforming the original network into an edge-centric network via the line graph. We discussed how directly operating on this construction leads to poor results and introduced an edge weighting mechanism for the line graph that preserves the dynamics of random walks. What also makes our approach useful is a combination of two aspects. First, it allows to revisit edge-centric tasks like community detection, that were introduced a decade ago, in the light of edge embeddings that were not consolidated at that time. Second, our approach also results useful in more recent downstream applications like path-based link prediction. 

We showed that embeddings paths as sequences of edges instead of sequences of nodes brings an immediate benefit. We also considered some potential limitations of our approach and proposed effective solutions. To face the fact that the line graph is much larger (in terms of nodes) than the original graph, we adopted a graph coarsening approach in the implementation. 

To deal with the potential larger number of parameters that results from the higher number of nodes in the line graph, we devised a strategy that automatically adapts the embedding dimension on the basis of the number of nodes of the original graph (the \systemD\ variant). There is room for several improvements. Considering alternative edge weighting mechanisms for \system\ and path aggregation strategies for \systemLP\ is in our research agenda.
%
%
\bibliographystyle{named}
\bibliography{biblio-embeddings}
\end{document}